\documentclass[pra,twocolumn]{revtex4-2}
\usepackage{amsmath}
\usepackage{amssymb}
\usepackage{amscd}
\usepackage{amsthm}
\usepackage{wasysym}
\usepackage[ansinew]{inputenc}
\usepackage[T1]{fontenc}
\usepackage{ae,aecompl}
\usepackage{amsfonts}
\usepackage{dsfont}
\usepackage{graphicx}
\usepackage{amsfonts}
\graphicspath{{./fig_paper/}}
\newcommand{\rmi}{{\rm i}}   
\newcommand{\rme}{{\rm e}}

\usepackage[caption=false]{subfig}
\usepackage{braket}
\usepackage{xcolor}
\usepackage{placeins}
\usepackage{printlen}
\newcommand{\Dfancy}{\mathcal{D}}
\newcommand{\Sfancy}{\mathcal{S}}

\begin{document}
	\title{Semiclassical Husimi distributions of Schur vectors in non-Hermitian quantum systems}
	\date{\today}
	\author{Joseph Hall}\author{Simon Malzard}\author{Eva-Maria Graefe}
	\address{Department of Mathematics, Imperial College London, London, SW7 2AZ, United Kingdom}
	\begin{abstract} 
		We construct a semiclassical phase-space density of Schur vectors in non-Hermitian quantum systems. Each Schur vector is associated to a single Planck cell. The Schur states are organised according to a classical norm landscape on phase space - a classical manifestation of the lifetimes which are characteristic of non-Hermitian systems. To demonstrate the generality of this construction we apply it to a highly non-trivial example a  PT-symmetric kicked rotor in the regimes of mixed and chaotic classical dynamics.
	\end{abstract}
	\maketitle

	
	
	The correspondence between quantum and classical mechanics has been investigated since the early days of quantum theory. For decades the main focus has been on quantum systems isolated from their environment, described by unitary time evolution. Most realistic systems, however, are subject to losses and dissipation, which can often be modelled by non-Hermitian Hamiltonians and non-unitary time evolution operators \cite{Moiseyev2011non,ashida2020non}.  The unique dynamical features of such systems have brought them into the spotlight, and there has been much effort to engineer specific non-Hermitian systems for the purpose of control of quantum and other wave dynamics \cite{el2018non,christodoulides2018parity}. The special class of PT-symmetric non-Hermitian Hamiltonians can lead to purely real eigenvalues and special dynamical properties \cite{Bender_1998_PT0,roberts1992chaos,bender2019pt}. The question on how insights into quantum-classical correspondence and semiclassical techniques can be extended to non-unitary quantum systems has come into focus in recent years. While considerable progress has been made for simple types of open quantum maps, modelling (partial) escape through openings \cite{Keating_2006,Hen_2010,Ketz_2018,Clauss2019,Ketz_2020,Ketz_2021_intensity_stats,Ketzmerick_2022,montes2023average}, the general question remains challenging.
	
	A common approach is the study of the localisation of quantum eigenstates on regions of classical phase space. The Husimi distribution of quantum states, which provides a quasi-probability distribution in phase space, is particularly useful in this context. In closed systems, the Husimi distributions of individual eigenstates localise on classical phase-space structures \cite{Berry_1977,Heller_1984,Bohigas_1993_manifestations}. A unit Planck cell supports a single eigenstate, i.e., to each state we may associate a distinct total area $h$ in phase space, and the sum of the Husimi functions of all eigenstates is the uniform distribution \cite{Bohigas_1993_manifestations,Hen_2012_weyl_law_open_sys}. This localisation lies at the heart of quantisation rules and algorithms for the construction of quasi modes. In systems described by non-normal operators the characteristic lack of orthogonality of the relevant quantum states presents a challenge in forming a similar correspondence. A consequence of the non-orthogonality is that a standard Planck cell partitioning of the phase space is no longer guaranteed and the sum of the Husimi distributions of all the eigenstates does not uniformly cover the whole phase space \cite{Hen_2010,Birchall_2012_Fractal_weyl}. 
	
In \cite{Hen_2004} it was argued for the case of a kicked rotor with escape through an opening in phase space, that considering the Schur vectors (arising from an orthogonalisation of the eigenvectors) allows to count phase space areas belonging to certain lifetimes consistently, leading to a fractal Weyl law \cite{lu2003fractal}.  In \cite{Hen_2010} the Husimi distributions of a set of Schur vectors was used for comparison to classical structures for a similar system. The underlying classical dynamics are not altered by the escape, but augmented with an additional lifetime. The support and localisation of ordered sets of Schur vectors were  heuristically found to correspond to regions of the phase space associated to trajectories with certain classical lifetimes. 

 In a similar vein in a study of a PT-symmetric kicked top a correspondence has been observed between Husimi distributions of Schur vectors and false colour plots of the semiclassical norm \cite{Eva_2020}. These studies point at a general connection between quantum Schur vectors and the behaviour of the classical counterpart of the norm as a function on phase space. The origin of this connection, and a quantitative description of the Husimi distribution of Schur vectors based on the semiclassical norm hitherto remain illusive. 

Here we argue that the correspondence of the Husimi distributions to areas of classical phase space belonging to certain values of the semiclassical norm can be traced back to the surprising observation that the eigenvectors of the Hermitian quantum operator $\hat W(t)=\hat U(t)\hat U^\dagger(t)$ for large times approach the Schur vectors of the time-evolution operator $\hat U$. Guided by this observation we exploit the orthogonality of Schur vectors to invoke the association of the support of individual states to unique Planck cells, that is familiar from Hermitian systems.  
This allows to systematically and uniformly cover the entire phase-space and recover quantisation rules. Here we devise an algorithm that constructs a classical Husimi density conditioned on the counting of  Planck cells. This generates a semiclassical approximation of the Husimi density of Schur states based solely on the classical dynamics. We demonstrate the construction for a PT-symmetric generalisations of a kicked rotor. 

	
Let us begin by reviewing some basic features of the eigenvectors and eigenvalues of non-Hermitian Hamiltonians. For simplicity we shall introduce these ideas for time-independent systems. They generalise to systems with periodic time-dependence. 

We consider a non-Hermitian Hamiltonian $\hat K$ with eigenstates $\ket{\phi_n}$,
\begin{equation}
\hat{K}\ket{\phi_n}=\epsilon_n\ket{\phi_n}
\end{equation}
where the energies $\epsilon_n=\varepsilon_n+i\mu_n$ are complex. The norm of an eigenstate changes in time as 
\begin{equation}
\braket{\phi_n|\hat U^\dagger(t)\hat U(t)|\phi_n}=\rme^{\frac{2\mu_n}{\hbar} t},
\end{equation}
where $\hat U(t)=\rme^{-\rmi\hat K t/\hbar}$ denotes the non-unitary time-evolution operator. Interpreting the norm as an overall probability, a gain-loss profile is encoded in the imaginary parts of the energy spectrum $\lbrace \mu_n\rbrace$  resulting in an exponential growth ($\mu_n>0$) or decay ($\mu_n<0$) of stationary states.  
	
In general, the eigenstates $|\phi_n\rangle$ are not orthogonal,  but away from exceptional points they form a complete basis for the relevant Hilbert space. Consider an arbitrary initial state 
$|\psi(t=0)\rangle$, which is expanded in the eigenbasis as
\begin{equation}
\ket{\psi(t=0)}=\sum_{n} \psi_n\ket{\phi_n},
\end{equation}
where $\psi_n=\langle\chi_n|\psi(t=0)\rangle$, and where $\lbrace\bra{\chi_n}\rbrace$ are the left eigenstates of $\hat K$ \cite{Brody2014}, normalised such that $\langle\chi_n|\phi_k\rangle=\delta_{nk}$. The state evolves as
\begin{equation}
\ket{\psi(t)}=\hat U(t)\ket{\psi(0)}=\sum_{n}e^{-\frac{i}{\hbar}\left(\varepsilon_n+i\mu_n\right)t}\psi_n\ket{\phi_n},
\end{equation}
which is similar to the unitary case, with additional relative exponential growth/decay of the coefficients of the different eigenstates. If we order the $\ket{\phi_n}$ such that 
\begin{equation}
\label{eqn:orderedgamma}
\mu_1>\mu_2>\ldots,
\end{equation}
an initial state that is a superposition of eigenstates will dynamically decay into the subspace spanned by the $|\phi_n\rangle$ with smaller $n$. Thus, there is a natural structure of invariant subspaces of the Hilbert space arising from the hierarchy of eigenvectors. To identify classical phase-space areas associated to these quantum subspaces, we need an orthogonal basis. This can be provided by appropriately ordered Schur vectors of the time-evolution operator, as suggested in \cite{Hen_2010}.
	
The non-Hermitian Hamiltonian can be expressed as $\hat{K}=\hat{V}\hat R\hat{V}^{\dagger}$ where $\hat{V}$ is a unitary matrix whose columns are the Schur vectors and $\hat R$ is an upper triangular matrix with the eigenvalues $\lbrace \epsilon_n  \rbrace$ on the diagonal. The Schur decomposition is not unique, but depends on the order of the eigenvalues along the diagonal. A natural choice is to order them according to growth/decay rates (\ref{eqn:orderedgamma}) reflecting the hierarchy of the quantum subspaces. It has been observed in two example systems that  sums of the Husimi functions of the Schur vectors belonging to the largest imaginary parts is localised on certain regions in classical phase space associated to these quantum subspaces \cite{Hen_2010,Eva_2020}.  
	
Here we identify these regions as bounded by contour lines of the classical counterpart of the norm in dependence of the initial state under propagation with $\hat U^\dagger$. This is connected to the surprising finding that in the (generic) case of complex quasi-energies with non-degenerate imaginary parts, the Schur vectors of $\hat K$, with the ordering (\ref{eqn:orderedgamma}), emerge as the asymptotic eigenvectors of the operator $\hat{W}(t)=\hat{U}(t)\hat{U}^{\dagger}(t)$ for large times. In Appendix \ref{Appendix:A} we motivate this observation with a concrete example and provide a general proof.
	
The operator $\hat W(t)$ is Hermitian, and thus, we can apply standard semiclassical arguments to associate phase-space areas bounded by contour lines of the classical phase-space function of the operator $\hat W$ with the quantum Husimi distributions of its eigenvectors, and by extension the Schur vectors of $\hat K$. The expectation value of $\hat W$ is the value of the norm of an initial state $|\psi(0)\rangle$ at time $t$, when we regard $\hat U^\dagger(t)$ as the time evolution operator. As a classical counterpart of this we identify the norm $w(q,p,t)$ of an initial coherent state centred at the respective phase-space point $(q,p)$ after propagation with $\hat U^\dagger(t)$ up to time $t$ under a coherent state approximation. Following \cite{Eva_2010}, this is found by integrating the dynamical equation 
\begin{equation}
\label{eqn:frozen_gaussian_norm}
\dot{w}(t)=2\Gamma(q(t),p(t))w(t),
\end{equation}	
along the phase space trajectories	
\begin{equation}
\label{eqn:frozen_gaussian_hamiltons_equations}
\dot{q}=-\frac{\partial H}{\partial p}+\frac{\partial \Gamma}{\partial q},
\ \ \
\dot{p}=\frac{\partial H}{\partial q}+\frac{\partial \Gamma}{\partial p}.
\end{equation}
where $H$ and $\Gamma$ denote the expectation values of the Hermitian and anti-Hermitian parts  $\hat H^\dagger=\hat H$ and $\hat \Gamma^\dagger=\hat \Gamma$ of the Hamiltonian $\hat K=\hat H+\rmi\hat \Gamma$.  

In addition to (potentially) modifying the phase-space trajectories, the gain/loss function $\Gamma$ introduces a non-trivial dynamics of the variable $w(t)$, the \textit{classical norm}. We represent the qualitative behaviour of the classical norm through false-color plots we refer to as \textit{norm landscapes}. The norm landscape is constructed by associating to each point $(q,p)$ of some distribution in phase space their respective average norm $\langle w(t) \rangle$ at a final time $t_f$. 

Based on the norm landscapes, we use a  state counting argument to construct a semiclassical approximation of the Husimi distributions of the quantum Schur vectors. We define the  subset $\Sfancy_m$ of phase space associated to the $m$-th  Schur state as the set of initial conditions that satisfy 
	\begin{equation}
		\label{eqn:defn_support_set_general}
		\Sfancy_{m}=\left\lbrace (q,p):\alpha_{-}^{(m)}\le \langle w_{t_f} \rangle\le\alpha_{+}^{(m)}\right\rbrace,
	\end{equation}
	where $\alpha_{-}^{(m)},\alpha_{+}^{(m)}\in\mathbb{R}$, and $\langle w_{t_f}\rangle$ is the average norm at a final time $t_f$. A priori the values $\alpha_{-}^{(m)}$, $\alpha_{+}^{(m)}$ are unknown, and are determined self-consistently from a semiclassical state counting argument. To mimic the intrinsic Gaussian nature of the Husimi distribution, we construct the classical density $\Dfancy_m(q,p)$ by the convolution of a uniform distribution on the set $\Sfancy_{m}$ with a two-dimensional Gaussian smoothing kernel with standard deviation $\sigma=\sqrt{\frac{\hbar}{2}}$ matching the minimum quantum uncertainty. We then condition the classical density by requiring it be normalised with respect to a Planck cell, that is, 
	\begin{equation}
		\label{eqn:husimi_norm}
		\int\int \Dfancy_{m}(q,p)dqdp=h.
	\end{equation}
		By simple extension, constructing the classical density for a group of $n$ Schur states, e.g. the $n$ states with largest $\mu$, simply requires the integral (\ref{eqn:husimi_norm}) to equal $nh$. 
  
  Let us now demonstrate these ideas for a PT-symmetric kicked rotor which features both chaotic and mixed classical dynamics, described by the  Hamiltonian
	\begin{equation}
	\label{eqn:ptkr_hamiltonian}
		\hat{K}=\frac{\hat{p}^2}{2}+\frac{i\gamma}{2\pi}\hat{p}+\frac{k}{4\pi^2}\cos(2\pi\hat{q})\sum_{n=-\infty}^{\infty}\delta(t-n),
	\end{equation}
	where $\gamma,k,\in\mathbb{R}^{+}$.  We consider the one-period evolution (Floquet) operator $\hat{U}$, where the time evolution is composed of the free evolution followed by a kick. We consider this example, because it yields a non-trivial classical dynamics, which differs from the Hermitian case in both the phase-space evolution and the additional norm. The example is in particular challenging due to non-trivial long-term dynamics brought about by the $PT$-symmetry of the system, in contrast to the behaviour of many other non-Hermitian or open systems, that approach limiting configurations on relatively short time scales. To demonstrate that our approach equally applies to open systems without PT-symmetry, in appendix \ref{Appendix:B} we apply it to a model of a  kicked rotor with partial escape that has been studied in \cite{Hen_2004,Hen_2010,Clauss2019}. 
 
 Using a standard quantisation on a torus \cite{Izrailev_1988_Torus_Model}, the matrix elements of the Floquet operator $\hat{U}$ for the model (\ref{eqn:ptkr_hamiltonian}) in the position representation are
	\begin{equation}
	U_{ll^{\prime}}\!=\!\frac{1}{N}e^{-\frac{iNk}{2\pi}\cos\left(\frac{2\pi l}{N}\right)}\!\!\!\sum_{m=-N_1}^{N_1}\!\!\! e^{-\frac{i\pi}{N}m^2+\gamma m + \frac{2\pi i }{N}m(l-l^\prime)},
	\end{equation}
	where $N=2N_1+1$ is the matrix dimension, which plays the role of an effective $\hbar=\frac{1}{2\pi N}$. The PT-symmetry manifests in the quasienergies $\epsilon_n$ being real or appearing in complex conjugate pairs.  The growth or decay rate is encoded in the imaginary part of a quasienergy $\mu_n=\text{Im}(\epsilon_n)$. Here we consider a matrix dimension of $N=1001$. For this value of $N$ even for small values of $\gamma$ for a broad range of kicking strengths $k$ a large number of quasi energies lie off the real axis in complex conjugate pairs. Spectral instabilities (which typically occur for nontrivial non-normal operators \cite{embree2005spectra}) limit the numerical range of $\gamma$ for which reliable quantum results can be obtained, and thus we will limit the discussions to small values of $\gamma\le0.003$. (The classical system can be analysed for any value of $\gamma$.)  
	
	The classical map generated by $\hat U^\dagger$ is given by 
	\begin{align}
		\nonumber	p_{n+1}&=\mod\left(p_n-\frac{k}{2\pi}\sin(2\pi q_{n})+\frac{\gamma}{2\pi}+\frac{1}{2},1\right)-\frac{1}{2}\\
		\label{eqn:ptkr_map_classical_bwds}	q_{n+1}&=\mod\left(q_n-p_{n+1}+\frac{\gamma}{2\pi},1\right).
	\end{align}
	While arising from a non-unitary evolution the classical map is in fact area preserving, independent of the value of $\gamma$, a curiosity that is made possible by the presence of PT-symmetry, but that is not necessarily typical for PT-symmetric systems. This area conservation prevents the formation of typical features of dissipative classical systems such as attractors, and chaotic saddles \cite{Altmann_2013_Leaking,Lai_2011_Transient}, yet, as we shall see, the quantum Schur vectors are clearly influenced by the presence of the loss and gain profile. This is reflected in the classical norm map (\ref{eqn:frozen_gaussian_norm}) 
	\begin{equation}
		\label{eqn:ptkr_map_norm}
		w_{n+1}=e^{2\gamma p_{n+1}}w_n.
	\end{equation}
	
	\begin{figure}
		\centering
		\subfloat{\includegraphics{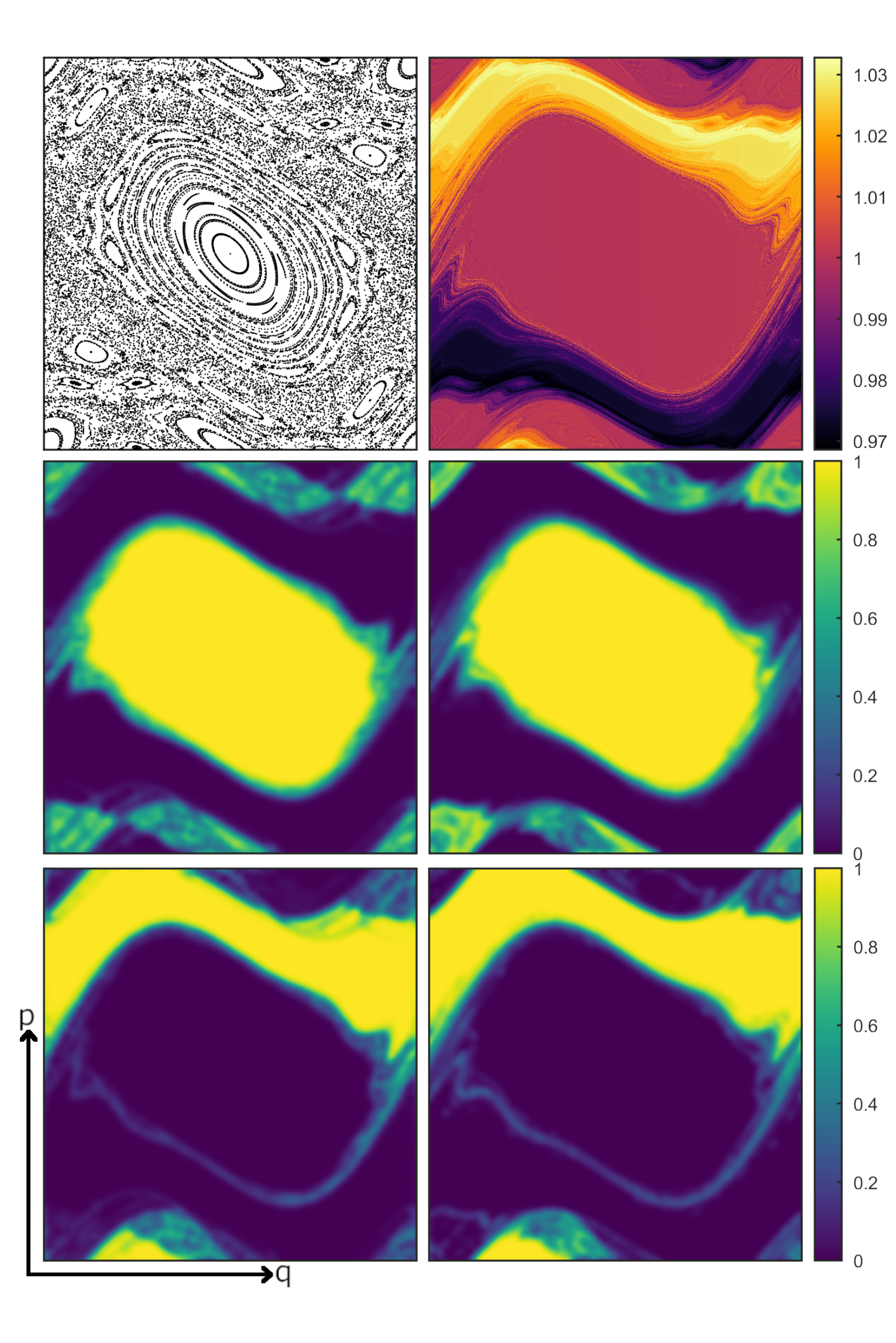}}
		\caption{Quantum-classical correspondence in phase space $(q,p)\in[0,1)\times[-0.5,0.5)$ for a non-Hermitian kicked rotor for  $k=1.1$ and $\gamma=0.001$. The top row depicts the Poincar\'e section (left) and the norm landscape (right) for a final time $t_f=66$. The middle and bottom rows depicts the Husimi distribution (left) of the stable (middle) and gain (bottom) set  and its associated classical density (right) for the same values of $k$, $\gamma$, $t_f$, and $\Delta_{+}=-\Delta_{-}=0.76$. }
		\label{fig:ptkr_qcc_mixed_full}
	\end{figure}
\begin{figure}
		\centering
		\subfloat{\includegraphics{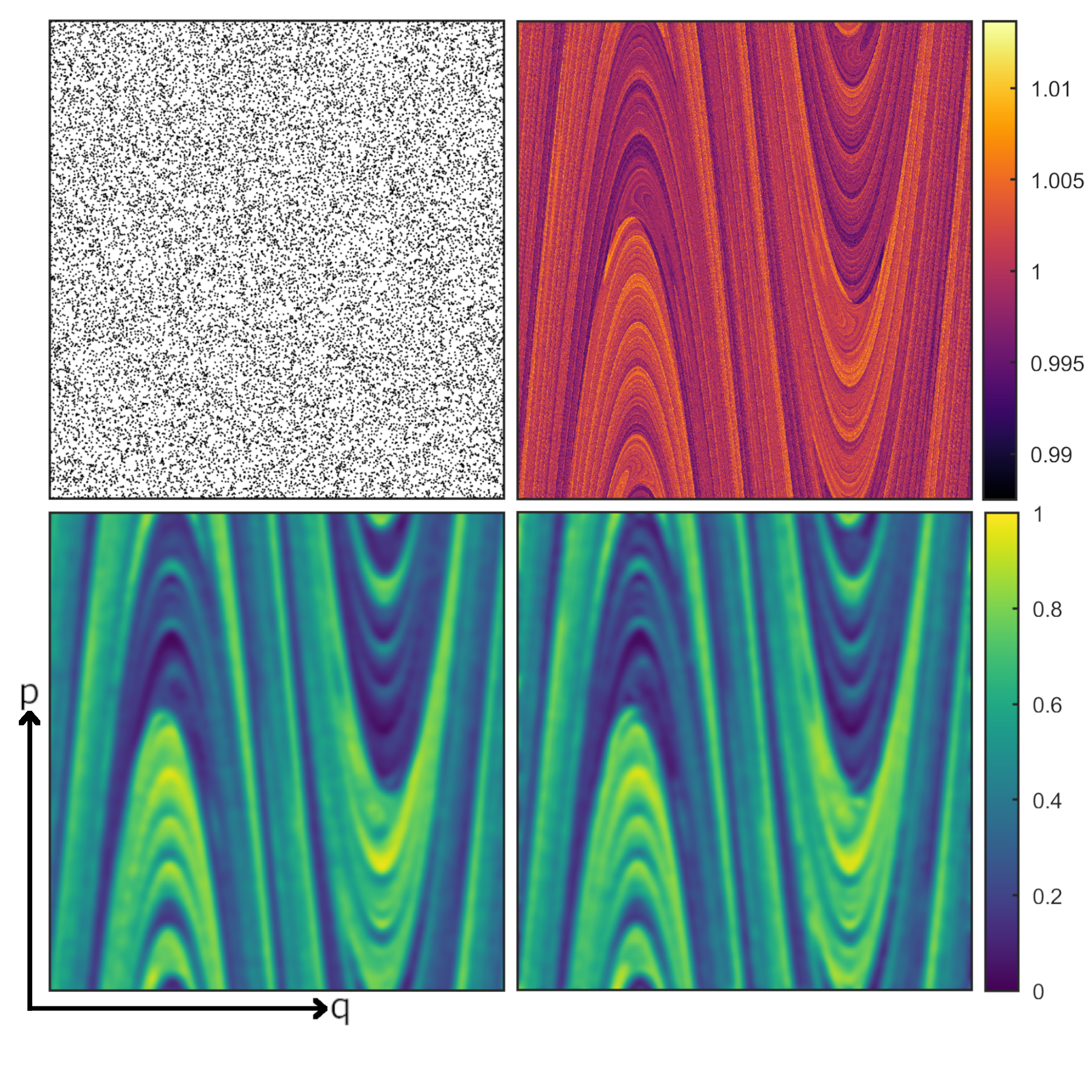}}
		\caption{Quantum-classical correspondence in phase space $(q,p)\in[0,1)\times[-0.5,0.5)$ for a non-Hermitian kicked rotor with $k=10$ and $\gamma=0.003$. From top left to bottom right:  Classical Poincar\'e section, norm landscape ($t_f=14$), Husimi distributions of the gain set and semiclassical phase-space density $\Dfancy_{+}$ for $t_f=14$ and $\Delta_{+}=0.048$.}
		\label{fig:ptkr_qcc_chaotic_full}
	\end{figure}

	In the left plot of the first row of Figure \ref{fig:ptkr_qcc_mixed_full} we show an example of a mixed regular-chaotic dynamics for $k=1.1$ and $\gamma=0.001$. For this small value of $\gamma$ the classical phase space trajectories differ little from those of the unitary system. The associated norm landscape for  $t=66$ is depicted in the right column of the same figure. We observe a rather distinct partitioning of the phase space into those trajectories whose norm grows, is stable, or decays. These regions can contain trajectories that are regular or chaotic, as can be seen in the regions of gain and loss that contain chains of regular islands. A more captivating example can be seen in the top row of Figure \ref{fig:ptkr_qcc_chaotic_full} where the Poincar\'e section (left) and norm landscape (right) are plotted for $k=10$, $\gamma=0.003$. Here the Poincar\'e section is dominated by a single featureless chaotic sea. In contrast, the norm landscape presents a much richer structure within the region corresponding to the chaotic sea in the Poincar\'e section. 
	
	The emergence of three different regions of norm behaviour is a classical manifestation of the PT-symmetry. The organisation of the Schur vectors around the norm landscape is demonstrated most readily in PT-symmetric systems by considering the Husimi distributions of three sets which we will refer to as gain, stable, and loss sets. The gain (loss) sets consist of those Schur vectors with positive (negative) imaginary parts of the quasienergy, $\mu_n>0$ ($\mu_n<0$) and the stable set corresponds to $\mu_n=0$. The left column of Figure \ref{fig:ptkr_qcc_mixed_full} depicts the Husimi distribution of the gain (middle row) and stable (bottom row) sets of Schur vectors for the case  $k=1.1$ and  $\gamma=0.001$. We observe a clear correspondence between the Husimi distributions of the gain and stable set and the relevant regions of the norm landscape. 
	Correspondingly, we define three classical support sets, $\Sfancy_{+}$ for the support on the gain set, $\Sfancy_{0}$ for the support on the stable set and $\Sfancy_{-}$ for the support on the loss set, as
	\begin{equation}
		\label{eqn:density_classical_sets_bwds}
		\begin{split}
			\Sfancy_{+}= \lbrace z &:w_{t_f}> e^{2\gamma\Delta_{+}}\rbrace\\
			\Sfancy_{-}= \lbrace z &:w_{t_f}< e^{2\gamma\Delta_{-}}\rbrace\\
			\Sfancy_{0}= \lbrace z&: e^{2\gamma\Delta_{-}}\le w_{t_f}\le e^{2\gamma\Delta_{+}}\rbrace.
		\end{split}
	\end{equation}
	The form $\alpha_{\pm}=e^{2\gamma\Delta_{\pm}}$ of the partitioning parameter is chosen to reflect the functional form of the classical norm (\ref{eqn:ptkr_map_norm}) and we have $\Delta_{\pm}\in\mathbb{R}^{\pm}$. Furthermore, the PT-symmetry of the system enforces the additional constraint $\Delta_{-}=-\Delta_{+}$.  Denoting the fraction of gain/loss states by $f_Q^{(\pm)}$ and the fraction of stable states by $f_Q^{(0)}$, we condition the corresponding classical densities $\Dfancy_{\pm,0}(q,p)$ such that 	
	\begin{equation}
		\label{eqn:Density_Condition}
		f_C^{(+)}=\frac{1}{N}\int\int \Dfancy_{+}(q,p) dqdp=f_Q^{(+)}, 
	\end{equation}
	and similarly for $f_Q^{(0)}$ and $f_Q^{(-)}$.
	
	The right column of Figure \ref{fig:ptkr_qcc_mixed_full} shows the classical densities associated to the two sets of Schur vectors in the left column, where the final times has been chosen as $t_f=66$. We observe a striking agreement between the Husimi distributions of the Schur vectors and the classical density for each case. Similarly Figure \ref{fig:ptkr_qcc_chaotic_full} shows the Husimi distribution (left) of the gain states and the classical density with $t_f=14$ (right) for $k=10$ and $\gamma=0.001$. Here the Husimi distribution show a more intricate structure than for $k=1.1$, which is again accurately reproduced by the classical density. 

	There typically is a large region of final times yielding similar results. We have verified that the Jensen-Shannon divergence \cite{Briet_2009_JSD,Ketz_2021_intensity_stats} between the quantum and classical Husimi densities rapidly decreases with the final time and reaches a plateau region. We have found minima of the Jensen-Shannon divergence at $t_f\approx 66$, $t_f\approx 14$ for $k=1.1$ and $k=10$, respectively, which we have chosen for the construction of the classical densities in Figures \ref{fig:ptkr_qcc_mixed_full} and \ref{fig:ptkr_qcc_chaotic_full}. In both cases choosing a $t_f$ in a relatively large range around the minimum value causes little quantitative difference on the results. 
	
In summary, we have introduced a classical density that yields an accurate approximation for the Husimi distributions of quantum Schur states for non-Hermitian systems with complex energies, as demonstrated for a PT-symmetric kicked rotor. This is a highly non-trivial model, with both complex and real quasi energies, with mixed and fully chaotic classical dynamics. The semiclassical Husimi distributions are constructed by associating to each Schur vector an area $h$ in the phase space determined by bands of the classical norm landscape. This is based on the curious observation that the Schur vectors emerge as asymptotic eigenvectors of the operator $\hat{W}(t)=\hat{U}(t)\hat{U}^{\dagger}(t)$, associated to the classical norm. Interestingly, these norm landscapes have been found to be of crucial importance in the classical propagation of Husimi distributions in non-Hermitian systems recently \cite{holmes2022husimi}. The deeper understanding of the meaning of the operator $\hat W$, the application of the proposed semiclassical algorithm to a larger class of example systems, and the extension to quantisation conditions for eigenvalues will be the topics of future investigations. 

The authors thank Roland Ketzmerick, Konstantin Clau\ss, Henning Schomerus, and Hans-J\"urgen Korsch for stimulating discussions and useful comments, and acknowledge  support  from the European Research Council (ERC) under the European Union's Horizon 2020 research and innovation program (grant agreement No 758453), E.M.G and J.H. acknowledge
support from the Royal Society (Grant. No. URF\textbackslash R\textbackslash 201034). 


\appendix
\, \\[0.2cm]
In these appendices we (a) provide details of the claim that for long times the eigenvectors of the norm operator $\hat W(t)=\hat U(t)\hat H^\dagger(t)$ approach the appropriately ordered Schur vectors of the time-evolution operator $\hat U$; and  (b) apply our new algorithm for a semiclassical approximation of the Husimi distributions of quantum Schur vectors to a well-studied example of a kicked rotor with partial escape from a certain region in phase space.

\section{Schur vectors of the time evolution operator as eigenvectors of the long-time norm operator} 
\label{Appendix:A}

To motivate the claim, let us demonstrate how the eigenvectors of $\hat W$ asymptotically approach the Schur vectors of the Hamiltonian $\hat K$ for a non-trivial but analytically solvable angular momentum model described by the non-Hermitian Hamiltonian
\begin{equation}
\hat K=\hat J_x+\rmi\gamma\hat J_z,
\end{equation}
where $\hat J_x$ and $\hat J_z$ denote standard angular momentum operators, and $\gamma\in\mathbb{R}$. We consider the case $|\gamma|>1$, for which the spectrum is purely imaginary. The Hamiltonian $\hat K$ is an element of the algebra $sl(2)$ and the eigenvalues of $\hat{K}$, for an $(N+1)$-dimensional representation, are given by $\rmi m\lambda$ where $m$ runs from $-\frac{N}{2}$ to $\frac{N}{2}$ in integer steps and $\lambda=\sqrt{\gamma^2-1}$. We can bring $\hat K$ to Schur form via the transformation
\begin{equation}
\hat K=\rme^{-\rmi\phi\hat J_x}(\rmi\lambda\hat J_z+\hat J_+)\rme^{\rmi\phi\hat J_x},
\end{equation}
where $\phi = \arctan{(\frac{1}{\lambda})}$ and $\hat J_+=\hat J_x+\rmi\hat J_y$ denotes the raising operator.
In the standard basis of $\hat J_z$ the eigenvalues appear in order of decreasing imaginary parts along the diagonal of the upper triangular matrix $\rmi \lambda\hat J_z+\hat J_+$. The Schur vectors of $\hat K$ are given by 
\begin{equation}
|v_m\rangle=\rme^{-\rmi\phi\hat J_x}|m\rangle,
\label{eqn:SchurvecSU2}
\end{equation} 
where $|m\rangle$ denotes the eigenvectors of $\hat J_z$. 
	
The (non-unitary) time evolution operator is given by $\hat U=\rme^{-\rmi \hat K t}$, and is an element of the group $SL(2)$. The operator $\hat W(t)=\hat U(t)\hat U^\dagger(t)$ is also is an element of $SL(2)$, and can be rewritten as 
\begin{equation}
\hat W(t)=A(t)\rme^{\theta(t)(c_y(t)\hat J_y+c_z(t)\hat J_z)},
\end{equation}
with time-dependent parameters $A\in\mathds{C}$, $\theta\in\mathds{R}$, and $c_y,c_z\in\mathds{C}$ with $c_y^2+c_z^2=1$.
For short times this is approximated by
$\hat W\approx\rme^{-2\gamma\hat J_z t}$, 
which shares the eigenvectors of $\hat J_z$. For long times, on the other hand we find
\begin{equation}
\hat W\to\frac{\rmi \gamma}{\lambda^2}\, \rme^{\lambda t}\, \rme^{\frac{\theta}{\gamma}\left(\lambda\hat J_z-\hat J_y\right)},
\end{equation}
with $\theta={\rm arctanh}\left(\frac{\gamma^2}{\lambda^2}\right)$. That is, the eigenvectors of $\hat W$ in the long time limit are the same as the eigenvectors of the operator $\hat X=\lambda \hat J_z-\hat J_y$ appearing in the exponent. Now $\hat X$ is easily diagonalised in the $\hat J_z$ basis as
\begin{equation}
\hat X=\rme^{-\rmi\phi\hat J_x}\hat J_z\rme^{\rmi\phi\hat J_x}, \quad{\rm with}\quad \phi={\rm \arctan}\left(\tfrac{1}{\lambda}\right).
\end{equation}
We thus find that the eigenvectors of $\hat X$ indeed coincide with the Schur vectors (\ref{eqn:SchurvecSU2}) of $\hat K$, verifying the claim for this specific  example.

For a general prove, let us consider the Schur decomposition of $\hat U = \hat V \hat R \hat V^\dagger$,  where $\hat R$ is an upper triangular matrix with the eigenvalues  $\rme^{-\rmi \epsilon_n t}$ (where $\epsilon_n=\varepsilon_n+\rmi\mu_n$) along the diagonal, such that 
\begin{equation}
\mu_1>\mu_2>\ldots,
\end{equation}
 and $\hat V$ is a unitary matrix composed out of the Schur vectors of $\hat U$. On the other hand, we assume that $\hat U$ can be diagonalised by a similarity transform as $\hat U=\hat S\hat D\hat S^{-1}$, where $\hat D$ is diagonal and $\hat S$ is composed out of the (non-orthogonal) eigenvectors of $\hat U$. We can rewrite $W$ in the basis of Schur vectors of $\hat U$ as
$$
W(t)=\hat U\hat U^\dagger=\hat V \hat R\hat R^\dagger \hat V^\dagger.
$$
We will now consider the matrix $\hat R\hat R^\dagger$, and show that in the limit of large times it indeed has the standard basis as eigenvectors. 

Diagonalising $\hat R=\hat S_R\hat D\hat S_R^{-1}$, with $\hat S_R=\hat V^\dagger \hat S$ we have
\begin{equation}
\hat R\hat R^\dagger= \hat S_R \hat D\hat S^{-1}\hat S^\dagger \hat D^* \hat S_R^\dagger.
\end{equation}
From this expression (using that $\hat S_R$ is an upper triangular matrix) we find that the matrix elements of $\hat R\hat R^\dagger$ are of the form
\begin{equation}
(\hat R\hat R^\dagger)_{jk}=\sum_{l=j}^{N} \sum_{l^{\prime}=k}^{N} e^{(\mu_l+\mu_{l^{\prime}})t} a_{jk}^{(l,l')} e^{-i\left(\varepsilon_l-\varepsilon_{l^\prime}\right)t},
\end{equation} 
where the $a_{jk}$ are complex time-independent constants with $a_{jk}=a_{kj}^*$
For large values of $t$ each of the elements of $\hat R\hat R^\dagger$ is dominated by a single term
\begin{equation}
(\hat R\hat R^\dagger)_{jk}\sim	 c_{jk}e^{(\mu_j+\mu_k)t},
\end{equation}
where the $c_{jk}$ are complex bounded (oscillatory) functions of time with $c_{jk}=c_{kj}^*$. 

We can now see how the standard basis emerges as eigenvectors in an asymptotic way. Let us factor out the dominant term $e^{2\gamma_1t}$, and shift the diagonal by an arbitrary ($t$-independent) amount $\lambda$. Then all elements of the matrix $\hat R\hat R^\dagger$, apart from the diagonal elements exponentially approach zero, and we have in the asymptotic limit 
\begin{equation}
\lim_{t\to\infty}\hat R\hat R^\dagger\sim c_{11}\rme^{2\mu_2 t}\begin{pmatrix} 1+\lambda &0 &\cdots & 0\\
0 & \lambda & \cdots & 0\\ 
\vdots & \vdots &\ddots & \vdots\\
0 & 0& \cdots  & \lambda
\end{pmatrix}
\end{equation}
Thus, the first eigenvector is clearly the vector $|e_1\rangle$. Since the remaining eigenvalues are degenerate, however, the corresponding eigenvectors are not uniquely determined once the limit of $t\to\infty$ has been taken. Taking into account the finite time corrections singles out the standard basis in the large time limit.

To see this, let us first consider the case $N=3$.  In the generic case the next largest elements after the dominant $(1,1)$ term are the $(1,2)$ and $(2,1)$ elements, and for large times we can write
\begin{equation}
\lim_{t\to\infty}\hat R\hat R^\dagger\sim c_{11}\rme^{2\mu_2 t}\begin{pmatrix} 1+\lambda & z&0\\
z^* & \lambda & 0\\
0&0&\lambda
\end{pmatrix},
\end{equation}
where $z$ is a small parameter, that decreases with increasing time. The eigenvectors of this matrix are given by the three vectors
\begin{equation}
\label{eqn_perteig}
\phi_1\propto\begin{pmatrix}\delta\\z^*\\0\end{pmatrix},\quad \phi_2\propto\begin{pmatrix} -z^*\\ \delta\\0\end{pmatrix},\quad \phi_3\propto\begin{pmatrix}0\\0\\1\end{pmatrix},
\end{equation}
where $\delta=\frac{1}{2}(1+\sqrt{1+4|z|^2})$. In the limit of large times, for which $z\to 0$, we have $\delta\to 1$, and the eigenvectors (\ref{eqn_perteig}) clearly approach the standard basis. 

In the case of $N=4$ we further need to consider the $(1,3)$ and $(3,1)$ as well as the $(2,2)$ elements, leading again to eigenvectors that are small perturbations of the standard basis. An analogous consideration can be carried forward to the general case.

We can thus show that if all the eigenvalues of $\hat U$ have different values of $\mu_j$, in the limit $t\to\infty$ the eigenvectors of the operator $\hat{W}(t)=\hat{U}(t)\hat{U}^{\dagger}(t)$ are indeed given by the Schur vectors of $\hat U$ in the appropriate ordering. 

\section{Kicked rotor with partial escape}
\label{Appendix:B}

\begin{figure}
\centering
\includegraphics{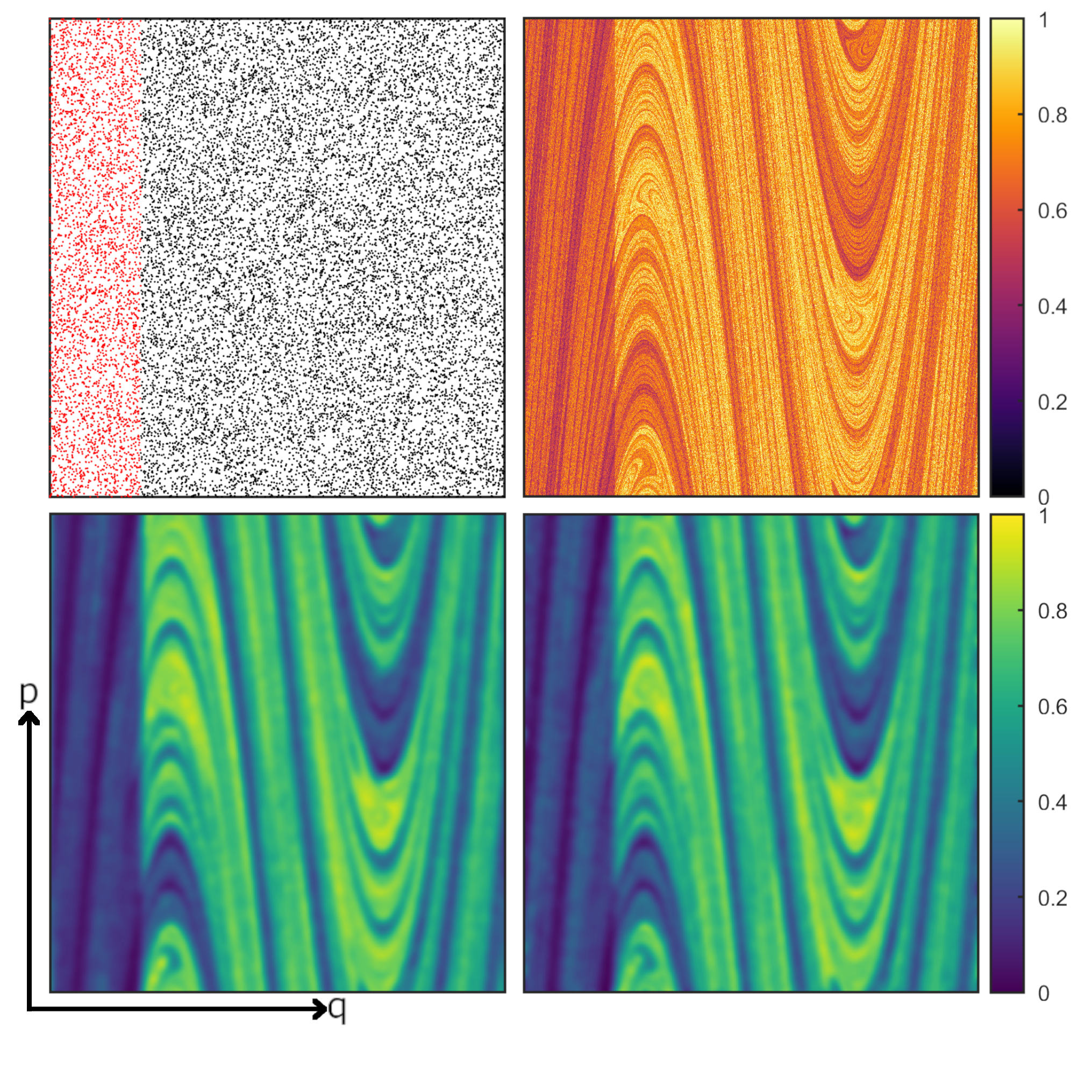}
\caption{Quantum-classical correspondence in phase space $(q,p)\in[0,1)\times[-0.5,0.5)$ for a lossy kicked rotor for $k=10$ and $\gamma=0.1$. The top row depicts the Poincar\'e section of the kicked rotor (left) with the region of loss $(q,p)\in[0,0.2)\times[-0.5,0.5)$ depicted in red and the norm landscape (right) for a final time $t_f=18$.  The bottom row depicts the Husimi distribution (left) of the set of $499$ Schur states with largest imaginary parts and the associated classical density (right) for a final time $t_f=18$ and partitioning parameter $\Delta_{499}=1.68$.}
\label{fig:nhkr_k10_full}
\end{figure}

To complement the example of a $PT$-symmetric kicked rotor presented in the main text, let us here consider a simpler model of a kicked rotor with partial escape described by the non-Hermitian Hamiltonian 
\begin{equation}
\label{eqn:nhkr_hamiltonian}
\hat{K}=\frac{\hat{p}^2}{2}-\frac{i\gamma}{4\pi^2}\chi(2\pi\hat{q})+\frac{k}{4\pi^2}\cos(2\pi\hat{q})\sum_{n=-\infty}^{\infty}\delta(t-n).
\end{equation}  
where $\gamma,k,\in\mathbb{R}^{+}$ and $\chi(q)$ is the characteristic function on the interval $q\in(q_L,q_R)$. The anti-Hermitian term induces a constant-loss, with strength proportional to $\gamma$, over a finite region of the phase space. This model is not $PT$ symmetric. It has been considered in \cite{Clauss2019}, and is closely related to the open systems considered in \cite{Hen_2004,Hen_2010,Ketz_2018}. We construct the quantum map in the same manner as for the example in the main text, where now the matrix elements in the position representation are given by
\begin{equation}
U_{ll^{\prime}}\!=\!\frac{1}{N}e^{-\frac{iNk}{2\pi}\cos\left(\frac{2\pi l}{N}\right)}\!\!\!\sum_{m=-N_1}^{N_1}\!\!\!\! e^{-\frac{i\pi}{N}m^2 + \frac{2\pi i }{N}m(l-l^\prime)-\frac{\gamma N}{8\pi^2}\chi\left(\frac{2\pi l}{N}\right)}
\end{equation}
 For non-zero $\gamma$ the quasienergies $\epsilon_n\in\mathbb{C}$ and their respective imaginary parts $\mu_n=\text{Im}(\epsilon_n)$ encode the decay rates. The classical map generated by $\hat{U^{\dagger}}$ is unchanged from the closed system and given by 
 \begin{align}
		\nonumber	p_{n+1}&=\!\!\!\!\mod\left(p_n-\frac{k}{2\pi}\sin(2\pi q_{n})+\frac{1}{2},1\right)-\frac{1}{2}\\
		\label{eqn:ptkr_map_classical_bwds}	q_{n+1}&=\!\!\!\!\mod\left(q_n-p_{n+1},1\right).
	\end{align}
 The loss manifests classically through the semiclassical norm which evolves as 
\begin{equation}
\label{eqn:nhkr_map_norm}
w_{n+1}=e^{-2\gamma\chi(q_n)}w_n.
\end{equation}
The top left plot of Figure \ref{fig:nhkr_k10_full} depicts the chaotic Poincar\'e dynamics for a classical kicking strength $k=10$. The points along the trajectories which lie inside the region of loss $0<q<0.2$ are depicted in red in the same figure. The influence of the loss on the system is reflected in the norm landscape which is depicted in the top right plot of Figure \ref{fig:nhkr_k10_full}, for  $\gamma=0.1$, a loss value which results in a non-trivial structure within the region of loss. Similar to \cite{Clauss2019,montes2023average}, we recognise the emergent structures in the norm landscape as belonging to the unstable manifold of the chaotic saddle, which for chaotic systems with partial escape is defined in terms of trajectories with non-zero intensity for arbitrary long times \cite{Altmann_2013_Leaking}. Here the norm landscape is shown for $t_f=20$ and thus shows features of both the unstable manifold of the saddle and the manifolds of transient trajectories that vanish asymptotically for longer times. The bottom left plot depicts the Husimi distribution of the set of $499$ Schur vectors of $\hat{U}$ with largest imaginary parts for a matrix dimension $N=1001$. The size of the set means that the Schur vectors cover nearly half of the total localisation area in the phase space and we observe localisation in both dynamical regions of the phase space. To construct the associated classical density we begin by defining the support set
\begin{equation}
\begin{split}
\Sfancy_{1:499}= \lbrace z &:w_{t_f}\ge e^{-2\gamma\Delta_{499}}\rbrace.
\end{split}
\end{equation}
The semiclassical density is then constructed by smoothing the set $\Sfancy_{1:499}$ with the Gaussian smoothing kernel and conditioning with the quantisation law such that
\begin{equation}
\label{eqn:Density_Condition_nhkr}
f_C^{(1:499)}=\frac{1}{N}\int\int \Dfancy_{1:499}(q,p) dqdp=f_Q^{(499)}.
\end{equation}
The resulting semiclassical density is depicted in the bottom row of Figure \ref{fig:nhkr_k10_full}, showing an excellent agreement with the quantum Husimi distribution. 

\end{document}